\documentclass[conference]{IEEEtran}
\IEEEoverridecommandlockouts
\usepackage{cite}
\usepackage{amsmath,amssymb,amsfonts}
\usepackage{algorithmic}
\usepackage{graphicx}
\usepackage{textcomp}
\usepackage{xcolor}
\usepackage{graphicx}
\usepackage{float}
\usepackage{subfig}
\usepackage{overpic}
\usepackage{comment}
\usepackage{url}

\def\BibTeX{{\rm B\kern-.05em{\sc i\kern-.025em b}\kern-.08em
		T\kern-.1667em\lower.7ex\hbox{E}\kern-.125emX}}
\begin{document}
	
	\newtheorem{definition}{\it Definition}
	\newtheorem{theorem}{\bf Theorem}
	\newtheorem{lemma}{\it Lemma}
	\newtheorem{corollary}{\it Corollary}
	\newtheorem{remark}{\it Remark}
	\newtheorem{example}{\it Example}
	\newtheorem{case}{\bf Case Study}
	\newtheorem{assumption}{\it Assumption}
	\newtheorem{property}{\it Property}
	\newtheorem{proposition}{\it Proposition}
	\newtheorem{observation}{\it Observation}
	
	\newcommand{\hP}[1]{{\boldsymbol h}_{{#1}{\bullet}}}
	\newcommand{\hS}[1]{{\boldsymbol h}_{{\bullet}{#1}}}
	
	\newcommand{\ba}{\boldsymbol{a}}
	\newcommand{\baq}{\overline{q}}
	\newcommand{\bA}{\boldsymbol{A}}
	\newcommand{\bb}{\boldsymbol{b}}
	\newcommand{\bB}{\boldsymbol{B}}
	\newcommand{\bc}{\boldsymbol{c}}
	\newcommand{\bcO}{\boldsymbol{\cal O}}
	\newcommand{\bh}{\boldsymbol{h}}
	\newcommand{\bH}{\boldsymbol{H}}
	\newcommand{\bl}{\boldsymbol{l}}
	\newcommand{\bm}{\boldsymbol{m}}
	\newcommand{\bn}{\boldsymbol{n}}
	\newcommand{\bo}{\boldsymbol{o}}
	\newcommand{\bO}{\boldsymbol{O}}
	\newcommand{\bp}{\boldsymbol{p}}
	\newcommand{\bq}{\boldsymbol{q}}
	\newcommand{\bR}{\boldsymbol{R}}
	\newcommand{\bs}{\boldsymbol{s}}
	\newcommand{\bS}{\boldsymbol{S}}
	\newcommand{\bT}{\boldsymbol{T}}
	\newcommand{\bw}{\boldsymbol{w}}
	
	\newcommand{\balpha}{\boldsymbol{\alpha}}
	\newcommand{\bbeta}{\boldsymbol{\beta}}
	\newcommand{\bOmega}{\boldsymbol{\Omega}}
	\newcommand{\bTheta}{\boldsymbol{\Theta}}
	\newcommand{\bphi}{\boldsymbol{\phi}}
	\newcommand{\btheta}{\boldsymbol{\theta}}
	\newcommand{\bvarpi}{\boldsymbol{\varpi}}
	\newcommand{\bpi}{\boldsymbol{\pi}}
	\newcommand{\bpsi}{\boldsymbol{\psi}}
	\newcommand{\bxi}{\boldsymbol{\xi}}
	\newcommand{\bx}{\boldsymbol{x}}
	\newcommand{\by}{\boldsymbol{y}}
	
	\newcommand{\cA}{{\cal A}}
	\newcommand{\bcA}{\boldsymbol{\cal A}}
	\newcommand{\cB}{{\cal B}}
	\newcommand{\cE}{{\cal E}}
	\newcommand{\cG}{{\cal G}}
	\newcommand{\cH}{{\cal H}}
	\newcommand{\bcH}{\boldsymbol {\cal H}}
	\newcommand{\cK}{{\cal K}}
	\newcommand{\cO}{{\cal O}}
	\newcommand{\cR}{{\cal R}}
	\newcommand{\cS}{{\cal S}}
	\newcommand{\dcS}{\ddot{{\cal S}}}
	\newcommand{\ds}{\ddot{{s}}}
	\newcommand{\cT}{{\cal T}}
	\newcommand{\cU}{{\cal U}}
	\newcommand{\wt}[1]{\widetilde{#1}}

	\newcommand{\mA}{\mathbb{A}}
	\newcommand{\mE}{\mathbb{E}}
	\newcommand{\mG}{\mathbb{G}}
	\newcommand{\mR}{\mathbb{R}}
	\newcommand{\mS}{\mathbb{S}}
	\newcommand{\mU}{\mathbb{U}}
	\newcommand{\mV}{\mathbb{V}}
	\newcommand{\mW}{\mathbb{W}}
	
	\newcommand{\uq}{\underline{q}}
	\newcommand{\ubq}{\underline{\boldsymbol q}}
	
	\newcommand{\red}[1]{\textcolor[rgb]{1,0,0}{#1}}
	\newcommand{\gre}[1]{\textcolor[rgb]{0,1,0}{#1}}
	\newcommand{\blu}[1]{\textcolor[rgb]{0,0,1}{#1}}
	
	\title{Rate-Distortion-Perception Theory for Semantic Communication
	}
	
	\author{\IEEEauthorblockA{Jingxuan~Chai\IEEEauthorrefmark{1}, Yong~Xiao\IEEEauthorrefmark{2}\IEEEauthorrefmark{3}\IEEEauthorrefmark{4}, Guangming~Shi\IEEEauthorrefmark{3}\IEEEauthorrefmark{1}\IEEEauthorrefmark{4}, Walid Saad\IEEEauthorrefmark{5} 
			\IEEEauthorblockA{\IEEEauthorrefmark{1}School of Artificial Intelligence, Xidian University, Xi'an, China}
			\IEEEauthorblockA{\IEEEauthorrefmark{2}School of Elect. Inform. \& Commun., Huazhong Univ. of Science \& Technology, China}
			\IEEEauthorblockA{\IEEEauthorrefmark{3}Peng Cheng Laboratory, Shenzhen, China}
			\IEEEauthorblockA{\IEEEauthorrefmark{4}Pazhou Laboratory (Huangpu), Guangzhou, China}
			\IEEEauthorblockA{\IEEEauthorrefmark{5}Bradley Department of Electrical and Computer Engineering, Virginia Tech, VA, USA}
		}\thanks{*This paper is accepted at IEEE International Conference on Network Protocols (ICNP) Workshop, Reykjavik, Iceland, October 10-13, 2023.}
	}
	
	\maketitle
	
	\begin{abstract}
		Semantic communication has attracted significant interest recently due to its capability to meet the fast growing demand on user-defined and human-oriented communication services such as holographic communications, 
		eXtended reality (XR), and human-to-machine interactions. Unfortunately, recent study suggests that the traditional Shannon information theory, focusing mainly on delivering semantic-agnostic symbols, will not be sufficient to investigate the semantic-level perceptual quality of the recovered messages at the receiver. In this paper, we 
		study the achievable data rate of semantic communication under the symbol distortion and semantic perception constraints. Motivated by the fact that the semantic information generally involves rich intrinsic knowledge that cannot always be directly observed by the encoder, we  consider a semantic information source that can only be indirectly sensed by the encoder. Both encoder and decoder can access to various types of side information that may be closely related to the user's communication preference.
		We derive the achievable region that characterizes the tradeoff among the data rate, symbol distortion, and semantic perception, which is then theoretically proved to be achievable by a stochastic coding scheme. We derive a closed-form achievable rate for binary semantic information source under any given distortion and perception constraints. We observe that there exists cases that the receiver can directly infer the semantic information source satisfying certain distortion and perception constraints without requiring any data communication from the transmitter. 
		Experimental results based on the image semantic source signal have been presented to verify our theoretical observations.

	\end{abstract}
	
	\begin{IEEEkeywords}
		Semantic communication, rate-perception-distortion, side information.
	\end{IEEEkeywords}
	
	\section{Introduction}
	
	Most existing communication systems are built mainly based on Shannon information theory which focuses on accurately delivering a set of semantic-agnostic symbols from one point to another. As mentioned in his classic work\cite{shannon1948mathematical}, Shannon believed that ``semantic aspects of messages" can be closely related to the users' background and personal preference, and therefore to develop a general theory of communication, it is critical to assume the ``semantic aspects of communication are irrelevant to the engineering problem".
	
	Recent development in mobile technology has witnessed a surge in the demand of user-defined and human-oriented communication services such as holographic communications, 
	eXtended reality (XR), and human-to-machine interactions, 
	most of which focusing on understanding and communicating the semantic meaning of messages based on users' background and preference. 
	This motivates a new communication paradigm, referred to as the {\em semantic communication}, which 
	focuses on transporting and delivering the meaning of the message.
	
	The concept of semantic communication was first introduced by Weaver in \cite{weaver1953recent} where it claims that one of the key differences between the semantic communication problem and Shannon's engineering problem is that, in the former problem, the main objective of the receiver is to recover a message that can ``match the \textit{statistical semantic characteristics} of the message" sent by the transmitter\cite{weaver1953recent}. In other words, the symbol-level distortion measures commonly used in the Shannon theory, e.g., Hamming distance, will not be sufficient to evaluate the semantic-level distortion of messages recovered by the receiver. In fact, a recent study has suggested that minimizing the symbol-level distortion does not always result in maximized human-oriented perceptual quality, commonly measured by the divergence between probability distributions of messages. 
	
	To address the above challenge, in this paper, we extend the Shannon's rate-distortion theory by taking into consideration of the constraint on the perceptual quality of semantic information.
	More specifically, we investigate the correlation between the symbol-level signal distortion and the semantic-level perception divergence as well as its impact to the achievable  data rate in a semantic communication system. Motivated by the fact that the semantic information generally involves rich intrinsic information and knowledge that cannot always be directly observed by the encoder, we consider a semantic information source that can only be partially or indirectly observed by the encoder. Also, both encoder and decoder can have access to various types of side information that can be closely related to the communication preference of the user as well as the indirect observation obtained by the encoder.
	We derive the achievable rate region that characterizes the tradeoff among the data rate, symbol-level signal distortion, and the semantic-level information perception, we referred to as the {\em rate-distortion-perception tradeoff}.  We prove that
	there exists a stochastic coding scheme that can achieve the derived tradeoff. We provide a closed-form achievable rate solution for the binary semantic information source under various distortion and perception constraints. We observe that there exist cases that the receiver can directly infer the semantic information under a certain distortion and perception limit without requiring any data communication  from the transmitter. 
	Experimental results based on the image-based semantic source signal have been presented to verify our theoretical observations.

	\section{Related Work}
	\subsection{Semantic Communications}
	Most existing works are focusing on directly extending the Shannon's classic information theory to investigate the semantic communication problem
	\cite{carnap1952outline,shi2021semantic,xiao2022imitation,xiao2023reasoning}. 
	For example, Carnap and Bar-Hillel replace the set of binary symbols of Shannon information theory with the set of possible models of worlds \cite{carnap1952outline}.
	The authors of \cite{bao2011towards} define the semantic entropy by replacing the traditional Shannon entropy of messages with the entropy of models.
	These earlier works on semantic communication always suffer from the Bar-Hillel-Carnap (BHC) paradox which states the self-contradictory or semantic-false messages often contain much more information, even they have less or no meaningful information, compared to basic facts or common sense knowledge.
	Recent works attempt to investigate the rate-distortion theory of semantic communication with indirect semantic source and multiple distortion constraints
	\cite{liu2021rate, xiao2022rate, stavrou2022rate}.
	Unfortunately, all of these works only consider the symbol-based distortion measure to evaluate the semantic distance, which is insufficient to 
	characterize the perceptual quality of semantic information recovered by the receiver. 
	In this paper, we take into consideration 
	the perception divergence between the real semantic information source and the semantic messages recovered by the receiver and investigate how various constraints in perception divergence can influence the achievable rate of the semantic communication.  
	
	\subsection{Rate-Distortion-Perception Tradeoff}
	Many efforts have been made recently to investigate the lossy source coding problem with  constraint on the perceptual quality. More specifically,
	Blau and Michaeli first formulate the information rate-distortion-perception function (RDPF) \cite{blau2018perception}.
	They further investigate the tradeoff of rate-distortion-perception on a Bernoulli source and MNIST dataset in \cite{blau2019rethinking}.
	Wagner establishes the coding theory of RDPF and proves its achievability \cite{theis2021coding}.
	Some recent works characterize the RDPF with side information and give some general theorems with respect to the achievability \cite{niu2023conditional,hamdi2023rate}.
	
	Additionally, recent works suggest that introducing common randomness at the encoder and decoder can be helpful for further improving perceptual quality. For example, Theis and Wagner suggest that stochastic encoders/decoders with common randomness as secondary input may be particularly useful in the regime of perfect perceptual quality by considering a 1-bit-coding toy example \cite{theis2021advantages}. Wagner further quantifies the amount of common randomness as one of the input of the RDPF with perfect perceptual quality, and shows the four-way tradeoff among rate, distortion, perception, and rate of common randomness \cite{wagner2022rate}.
	However, some other recent works suggest that the benefit  of common randomness is not always available.
	The authors of \cite{niu2023conditional} and \cite{chen2022rate} show that common randomness is not necessary under the empirical distribution constraints of perceptual quality.
	Motivated by these works, in this paper,  we consider a semantic communication system in which two types of side information, including the Wyner-Ziv side information and common randomness, can be available at the encoder and decoder. We characterize RDPF under both the empirical and strong perception constraints. We observe that under certain scenarios that the side information along is sufficient for the decoder to fully recover the semantic information source.   

	\section{Problem Formulation and Main Results}
	We consider the semantic communication model introduced in \cite{xiao2022rate}. In this system, a semantic information source involves some intrinsic knowledge or states that cannot be directly observed by the encoder. The encoder can collect a limited number of indirect observations, i.e., explicit signals that reflect some properties or attributes of the semantic information source. The encoder will then compress its indirect observations according to the channel capacity to be sent to the destination user. The main objective is to recover the complete information including all the intrinsic knowledge with minimized semantic dissimilarity, called semantic distance, at the destination user. More formally, the semantic information is formulated as an $n$-length i.i.d. random variable generated by the semantic information source.   
	The encoder can obtain a limited number $k$ of indirect observations about the semantic information source, i.e., the input of the encoder is given by  $X^k$. The indirect observation of the encoder may correspond to a limited number of random sampling of noisy observations about the semantic information source obtained by the encoder. We will give a more detailed discussion based on a specific type of semantic information source, e.g., binary source, in Section \ref{Section_BinarySource}.
	Let $p_{X|S}$  be the conditional probability of obtaining an indirect observation $X$ at the encoder under a given semantic information source $S$. 
	Due to the limited channel capacity, the encoder can only send $m$ sequences of signals, denoted as $M^m$, to the channel.
	
	In addition to observe the output of the channel, the decoder can also have access to two types of side information about the encoder and the semantic information source:
	
	\noindent{\bf Common Randomness (CR):}  
	corresponds to a source of randomness that is involved in the encoding function to generate the coded messages sent to the channel. This may correspond to stochastic encoder in which the input of encoder includes both indirect observation and a source of randomness. The source of randomness can be the result of a randomized quantization process included in the encoding process for compressing the indirect observation. It may also correspond to the randomness added to the encoded messages for improving the protection of privacy of the indirect observation, e.g., in differential privacy-enabled encoding. The randomness of the encoding process can be considered as a pseudo-random variable generated by a seed that is known by the decoder.
	We assume the randomness of the encoder can be represented by
	a real-valued random variable,
	denoted by $U$.
	
	\noindent{\bf Wyner-Ziv (WZ) Side Information:}
	corresponds to some background information related to the  indirect observation of the semantic information source. For example, in the human communication scenario\cite{xiao2022rate}, different human users may use different languages or preferred sequences of words to express the same idea. In this case, the WZ side information may correspond to the language background and preference of users when converting the semantic information source, i.e., ideas, to a sequence of words. Generally speaking, both encoder and decoder may have some side information. The side information at the encoder and decoder does not have to be the same. Let $Y'$ and $Y''$ be the WZ side information at the encoder and decoder, respectively.   
	
	
	Both encoder and decoder will include the WZ side information and CR in their coding process. More formally, we define the stochastic code as follows:
	\begin{definition}
		For an arbitrary set $\mathcal{X}^k$, a (stochastic) encoder is a function
		\begin{eqnarray}
			f:\mathcal{X}^k\times\mathcal{Y}'\times\mathbb{R}\to \{1,2,...,2^{mR}\},
		\end{eqnarray}
		and a (stochastic) decoder is a function
		\begin{eqnarray}
			g:\{1,2,...,2^{mR}\}\times\mathcal{Y}''\times\mathbb{R}\to \mathcal{S}^n.
		\end{eqnarray}
	\end{definition}	
	In this paper, we consider two types of semantic distances:

	\noindent{\bf Block-wise Semantic Distance: } corresponds to a block-wise distortion measure defined as a function $d: S^n\times {\hat S}^n \rightarrow \mathbb{R}$. In this paper, we consider the following block-wise semantic distance constraint: $\mathbb{E} [d(S^n, {\hat S}^n)] \le D$.
	
	\noindent{\bf Perception-based Semantic Distance: } corresponds to non-negative divergence between any given pair of probability distribution functions. In this paper, we adopt the total variance (TV) distance, a commonly used divergence function, to measure the perception-based semantic distance between the real semantic information source $S$ and the estimated signal $\hat S$ at the decoder, defined as follows:
	\begin{eqnarray}
		D_{TV}(p_S,p_{\hat S})=\sup_{A\in\mathcal{S}}|p_S(A)-p_{\hat S}(A)|.
	\end{eqnarray}
	
	
	Let us formally define achievable rate region of the rate-distortion-perception tradeoff as follows:
	\begin{definition}
		The rate-distortion-perception triple $(R,D,P)$ is achievable with respect to strong and empirical perception constraints if for any $\epsilon>0$, there exists an $(n,2^{n(R+\epsilon)})$
		code with common randomness $U\in\mathbb{R}$ that satisfies the direct bit distortion constraint
		\begin{eqnarray}
			\label{dh}
			\mathbb{E}[d(S,\hat S)]\leq  D+\epsilon,
		\end{eqnarray}
		and one of the indirect perception constraints		
		\begin{equation}
			\label{dtvs}
			\begin{split}
				D_{TV}(p_{S^n},p_{\hat S^n}) &\leq P+\epsilon,  \textrm{(for strong perception)} \\
				\mathbb{E}[D_{TV}(\hat p_{S^n},\hat p_{\hat S^n})] &\leq P+\epsilon, \textrm{(for empirical perception)}.
			\end{split}
		\end{equation}
		
		Note that the main difference between the strong and empirical perception constraints is that in the later case,  the TV distance is measured by expectations  between two probability  distributions, while for strong perception constraint, the TV distance is calculated based on the $n$-length code block.
	\end{definition}
	
	\begin{definition}
		The rate-distortion-perception region under strong perception constraint is
		\begin{eqnarray}
			\mathcal{R}^{(s)}=&\left\{\right.(R,D,P):\exists\ Z,\textrm{s.t.}\ p_{SXYZ}=p_Sp_{XY|S}p_{Z|X},\nonumber\\
			& R\geq I(X;Z|Y), I(S,X;Z|Y)\leq \frac{k}{m}I(M;\hat M),\nonumber\\
			&\mathbb{E} [d(S^n, {\hat S}^n)]\leq D,D_{TV}(p_{S^n},p_{\hat S^n})\leq P\left.\right\}.
		\end{eqnarray}
		And the region for empirical perception constraint, $\mathcal{R}^{(e)}$, is defined by replacing the perception constraint of above region with $D_{TV}(p_S,p_{\hat S})\leq P$.
	\end{definition}

	We can then prove the following result about the achievability of the rate-distortion-perception region.
	\begin{theorem}
		\label{th1}
		$(R,D,P)$  is achievable with respect to strong (resp. empirical) perception constraint if and only if it is contained in the closure of $\mathcal{R}^{(s)}$ (resp. $\mathcal{R}^{(e)}$).
	\end{theorem}
	\begin{IEEEproof}
		See Appendix \ref{proof_th1}.
	\end{IEEEproof}
	\begin{observation}
		Definition 3 specifies the constraint of achievable coding rate of our proposed semantic communication system. We can then reformulate the achievable rate $I(S,X;Z|Y)$ as following forms using the chain rules of mutual information,
		\begin{eqnarray}
			\label{mutual1}
			I(S,X;Z|Y)&=&I(S,X;Z,Y)-I(S,X;Y)\\
			&=&H(X|Y)+H(S|X,Y)+H(Z|Y) \nonumber\\
			\label{mutual2}
			& \;&-H(S,X,Z|Y).
		\end{eqnarray}
		The first term of the right-hand-side of (\ref{mutual1}) specifies the amount of uncertainty about the indirect semantic information $S$ and direct observation $X$ that can be reduced at the receiver.
		The second term of the right-hand-side of (\ref{mutual1}) specifies the total amount of uncertainty induced by the existence of side information, which is not required to be transmitted in the channel.
		(\ref{mutual2}) quantifies the amount of information of each individual source in terms of conditional entropy.
		Specifically,
		$H(X|Y)$ specifies the amount of information of the indirect semantic source.
		$H(S|X,Y)$ quantifies the semantic ambiguity about the semantic information, induced by indirect observation.
		$H(Z|Y)$ specifies the volume of information to be encoded at the receiver.
		$H(S,X,Z|Y)$ quantifies the amount of uncertainty about the semantic information and the channel output which can be reduced when using side information at the decoder.
		Moreover, as for the informative correlation between the semantic source and the indirect observation, we have
		$
		H(X)=H(S)+H(X|S)-H(S|X),
		$
		where $H(X|S)$ corresponds to the semantic redundancy and $H(S|X)$ corresponds to the semantic redundancy.
	\end{observation}
	
	\section{Closed-Form Achievable Rate of Binary Semantic Information Source}
	\label{Section_BinarySource}

	In this section, 
	we focus on the binary semantic information source scenario in which the semantic source follows a Bernoulli distribution $S\sim Ber(\pi)$ with $\pi\leq\frac12$. 

	The correlation between indirect observation and the semantic information source 
	and that between the side information adn semantic information source can be parameterized as:
	$
		p_{X|S}=
		\begin{bmatrix}
			1-q_1 & q_1 \\
			q_2 & 1-q_2
		\end{bmatrix},
		p_{Y|S}=
		\begin{bmatrix}
			1-u & u \\
			v & 1-v
		\end{bmatrix}.
	$
	
	We then move on to present the closed-form solution of RDPF under binary alphabet setting. To simplify the expression, we first parameterize the correlation between $X$ and $Y$ as following matrix:
	$
	p_{Y|X}=
	\begin{bmatrix}
		1-a & a \\
		b & 1-b
	\end{bmatrix}.
	$
	We also parameterize the conditional probabilities of semantic source $X$ and indirect observation $X$ given side information $Y$ as following matrices
	$
		p_{X|Y}=
		\begin{bmatrix}
			1-a^* & a^* \\
			b^* & 1-b^*
		\end{bmatrix},\
		p_{S|Y}=
		\begin{bmatrix}
			1-u^* & u^* \\
			v^* & 1-v^*
		\end{bmatrix}.
	$
	Then define the $\pi$-RDF:
	\begin{eqnarray}
		R_\pi(D):= H_b(\pi)-H_b(D),
	\end{eqnarray}
	and $\pi$-RDPF:
	\begin{equation}
		\begin{split}
			R_\pi(D,P):=&2H_b(\pi)+H_b(\pi-P)-H_t(\frac{D-P}{2},\pi)\\
			&-H_t(\frac{D+P}{2},1-\pi),
		\end{split}
	\end{equation}
	where $H_b$ denotes the entropy of a binary variable, $H_t$ denotes the entropy of a ternary variable.
	Then we have following theorem with respect to the closed-form expression of the rate-distortion-perception function $R(D,P)$:
	
	\begin{theorem}
		\label{th2}
		Assume $u^*=v^*<\frac12$, $a^*=b^*=\pi_x\leq \frac12$ and $X,S$ follow the doubly symmetric binary distribution with $q_1=q_2=q$,
		the achievable rate under distortion $D$ and $P$ equals to a rate-distortion-perception function, formulated as
		\begin{eqnarray}
			R(D,P)=
			\left\{
			\begin{aligned}
				R_{\pi_x}(D_x)\quad &{\rm if}\ D\in[q,\pi_x')\\
				0\hspace{24pt} &{\rm if}\ D\in [\pi_x',\infty),
			\end{aligned}
			\right.
		\end{eqnarray}
		for $P\geq \pi_x$ and
		\begin{eqnarray}
			R(D,P)=
			\left\{
			\begin{aligned}
				R_{\pi_x}(D_x),\hspace{24pt} &{\rm if}\ D\in[q,D')\\
				R_{\pi_x}(D_x,P),\hspace{16pt}&{\rm if}\ D\in[D',\pi_x')\\
				0,\hspace{40pt} &{\rm if}\ D,\in[\pi_x',\infty)\\
			\end{aligned}
			\right.
		\end{eqnarray}
		for $0<P<\pi_x'$,
		where $D_x=\frac{D-q}{1-2q}$,
		$\pi_x'=(1-2q)\pi_x+q$,
		$D'=\frac{(1-2q)P}{1+2P-2\pi_x}+q$.
	\end{theorem}
	\begin{IEEEproof}
		See Appendix \ref{proof_th2}.
	\end{IEEEproof}
	
\begin{figure} [htbp]
		\centering
		\subfloat[]{
			\includegraphics[scale=0.28]{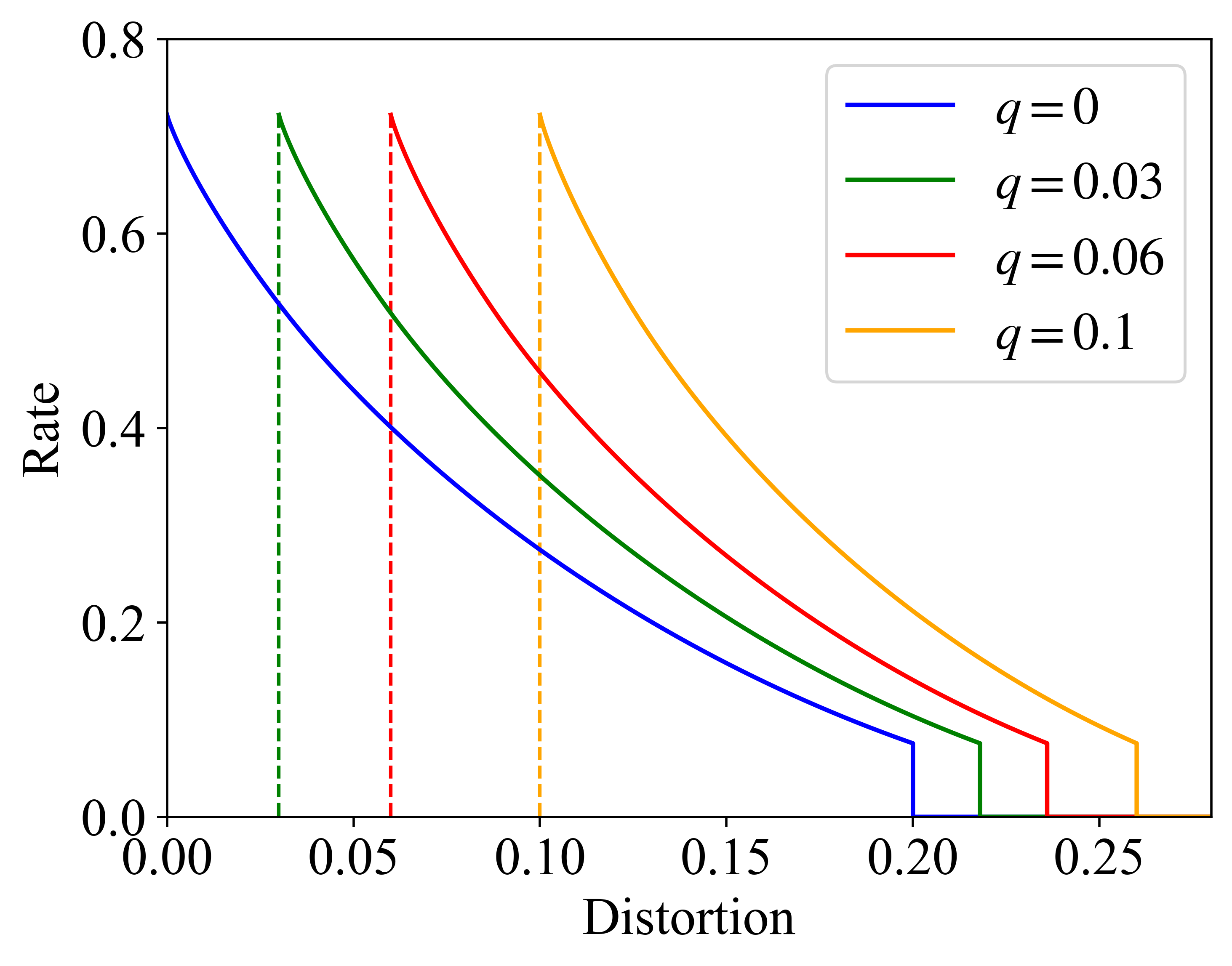}
			\label{1a}	
		}
		\subfloat[]{
			\includegraphics[scale=0.28]{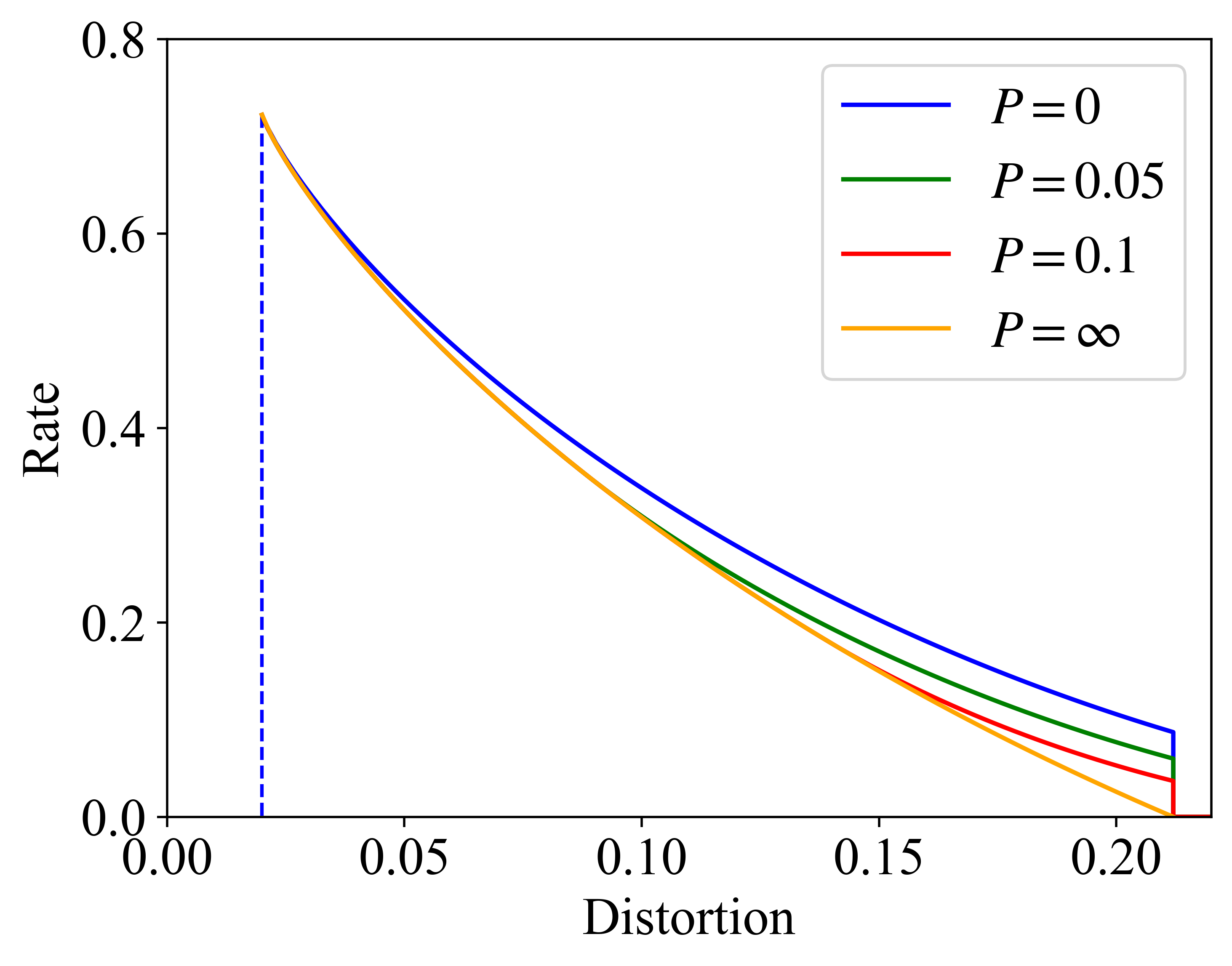}
			\label{1b}}
		\caption{RDPF of a binary semantic information source under various crossover probabilities and perception constraints.}
		\label{2figs}
	\end{figure}

	To evaluate the impact of crossover probabilities and constraints of perceptions on the achievable rate of semantic communication, we present RDPF results in Fig. \ref{2figs} where we set $a=b=0.2$. We have the following observations:  
	We then make several observations as follows.	
	\begin{observation}
		In Fig.\ref{2figs}, we can observe that higher data rate is required when the constraints on the symbol-level distortion and higher perceptual quality are more stringent.
		In particular, in Fig.\ref{1b}, under  a given distortion level,
		higher achievable rate is demanded when requiring higher perceptual quality, i.e., value of $P$ is lower.
		Similarly, under a given achievable rate,
		the perceptual quality decreases when requiring higher symbol-level quality.
		The above observations about the rate-distortion-perception tradeoff  is consistent with the traditional RDPF without side information \cite{blau2019rethinking}. 
	\end{observation}
	\begin{observation}
		In the proof of Theorem 2, we can observe that the distortion  between semantic information source and the recovered information information, denoted as $d(S, \hat S)$, and that between the direct observation obtained by the encoder and the the recovered information information, denoted as $d(X, \hat S)$, has the following relationship:
		\begin{eqnarray}
			d(X,\hat S)=({d(S,\hat S)-q})/({1-2q}). \label{eq_dXhatS}
		\end{eqnarray}
		We can observe from (\ref{eq_dXhatS}) that, since both $d(S, \hat S)$ and $d(X,\hat S)$ must be positive, the 
		symbol-level distortion $d(S, \hat S)$ cannot be lower than $q$.
		Also, when $d(X,\hat S) = 0$, we have $d(S,\hat S) = q$. This means that trying to perfectly recover the indirect observation obtained by the encoder cannot result in perfect recovery of the semantic information source at the decoder. This further justifies that the objective of the semantic communication which tries to recover the semantic information source, i.e., minimizing $d(S,\hat S)$, is generally different from that of the traditional communication solution which focusing on recovering the symbol obtained by the encoder, i.e., minimizing $d(X,\hat S)$.
	\end{observation}
	\begin{observation}
		In Fig.\ref{1a}, the blue curve represents the direct observation case with $q=0$ in which the encoder can fully observe the semantic information source, i.e., $X=S$. We can observe that the indirect observation of the semantic information source, i.e., $q>0$, leads to a higher required data rate under the same distortion and perception constraints, compared to the direct observation case.
		And the difference between rates achieved by the direct and indirect observations at the encoder increases with the value of $q$.
	\end{observation}
	\begin{observation}
		In Fig.\ref{2figs}, we can observe that when the rate $R(D, P)$ decreases to zero, the symbol-level distortion $D$ will not increase to the maximum value $0.5$, but will stick to a fixed value, given by $\pi_x$. 
		This means that when the side information is available at both encoder and decoder, the decoder is able to recover semantic information source under certain distortion constraints even when no data has been transmitted from the transmitter. More specifically, suppose $D\geq \pi_x$ and $\hat S=Y$. By substituting these two equations into (\ref{dtvs}), we have
		$
		D_{TV}(p_S,p_{\hat S})=D_{TV}(p_X,p_{\hat S})=D_{TV}(p_X,p_Y)
		=\frac12|a-b|=0.
		$
		This means the perception divergence becomes zero when $D\geq \pi_x$.
		The optimal decoding strategy in this case is then given by $\hat S=Y$.
		This implies that the decoder can directly use the side information to recover $S$ when $D\geq \pi_x$. This may correspond to the case that the end user can directly infer the semantic  information source based only on the background knowledge of source user as long as the required distortion and perception constraints are above a certain tolerable threshold.
	\end{observation}

	\section{Experimental Results}
	
	\subsection{Experimental Setup}
	We consider a semantic communication system in which the semantic information source  $S$ corresponds to image signals that are uniformly randomly sampled from a commonly used image dataset, MNIST.
	We simulate the indirect observation of the semantic source information by  adding a  noise signal to the semantic source, i.e. each indirect observation $x$ is given by
	$
	x=s+n_s,
	$
	where $s$ is the input image, $n_s$ is the noise vector, $x$ denotes the indirect observation obtained by the encoder.
	The side information is some intrinsic feature of semantic source, denoted as $y=f_Y(x)$, where $y$ denotes the side information of $x$, $f_Y$ denotes the function that output the feature of $x$, parameterized by deep neural nets (DNNs).
	The encoder maps the indirect observation $x$ into a $d-$dimensional latent feature vector, whose entries are then uniformly quantized to $L$ levels to obtain the output message.
	The decoder finally recovers the semantic information based on received message and the side information.
	Both the encoder and decoder have access to the common randomness, modeled as a Gaussian noise vector, denoted as $n_c$.
	The entire process of semantic source coding can be formulated as
	$
	\hat s=f_D(f_E(x,n_c),y,n_c),
	$
	where $f_E$ and $f_D$ denote the encoding and decoding function.
	Note that the uniform quantization of the encoder gives an upper bound of $d\log(L)$ for rate $R$. The quantization level is set to be $L=8$.
	
	We measure symbol-based signal distortion using the mean-squared-error (MSE), denoted as $\mathbb{E}(\|X-\hat S \|^2)$. We also use the Wasserstein distance $D_W(p_S,p_{\hat S})$ to measure the perception divergence between distributions $p_S$ and $p_{\hat S}$.
	We then train the our model to minimize following loss function,
	$
		\mathcal{L}=\mathbb{E}(||S-\hat S||^2)+\lambda D_W(p_S,p_{\hat S}),
	$
	where $\lambda$ is a tuning parameter which controls particular tradeoff point achieved by the model.
	Based on \cite{blau2019rethinking}, the above perceptual quality term can be written as
	$
		D_W(p_S,p_{\hat S})=\max_{h\in\mathcal{F}}\mathbb{E}h(S)-\mathbb{E}h(\hat S),
	$
	where $\mathcal{F}$ is the set of all bounded 1-Lipschitz functions.
	This expresses the objective as a min-max problem and allows us to treat it using the generative adversarial network (GAN).
	We obtain the experimental  RDPF curves by evaluating the distortion and perceptual quality of our model on the test set of MNIST. Different rate-distortion-perception points are obtained by training the model under different desired settings.
	
	\subsection{Experimental Results}
	\begin{figure}[htbp]
		\centering
		\includegraphics[width=0.27\textwidth]{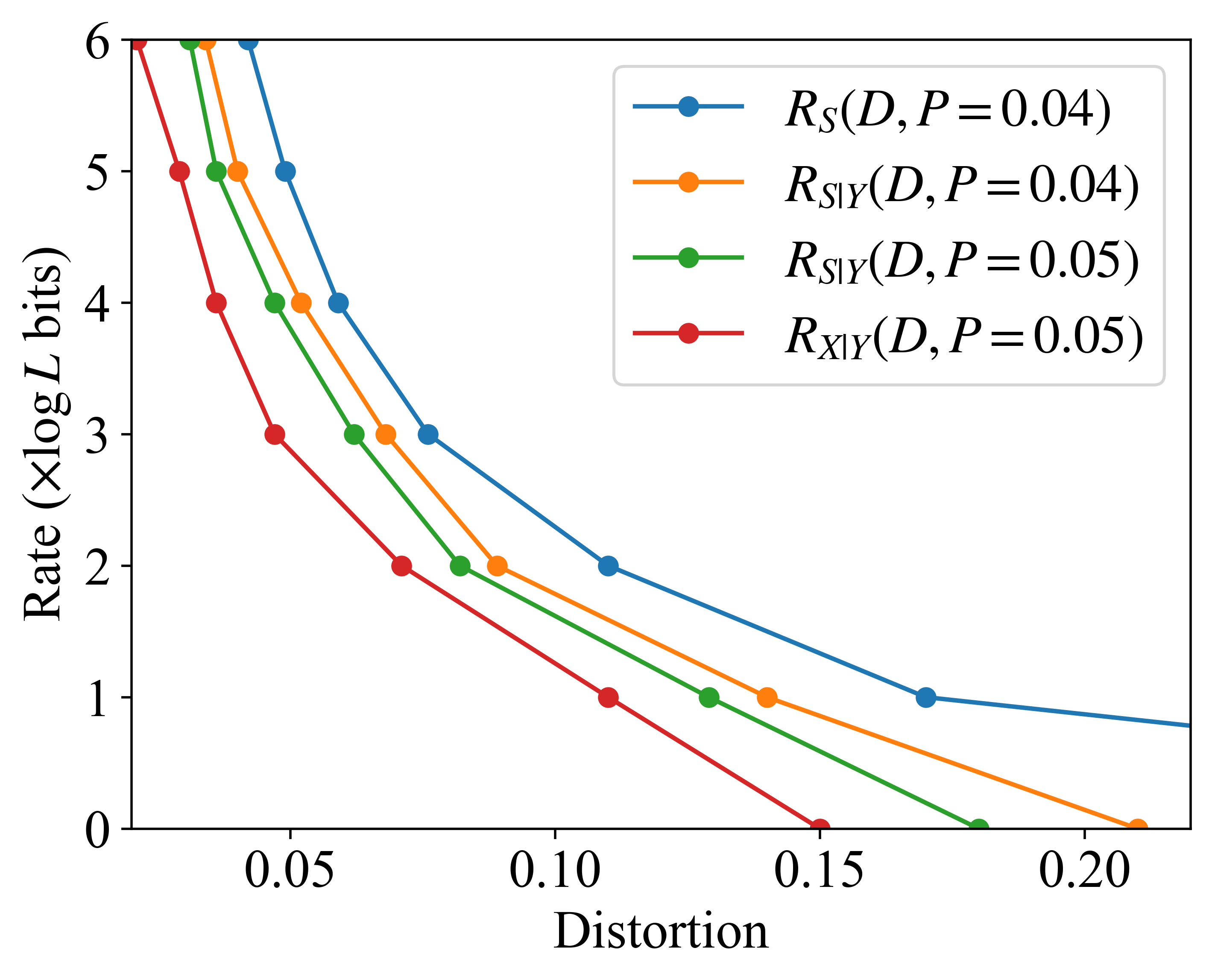}
		\caption{RDPF with and without side information $Y$ at the decoder when the semantic information source corresponds to image signals sampled from the MNIST dataset.}
		\label{exp}
	\end{figure}
	In Fig. \ref{exp}, we compare the RDPF curves with and without the side information $Y$ at the decoder which are denoted as $R_{S|Y}$ and $R_S$, respectively. We also present the RDPF curve when the encoder can obtain a full observation of the semantic information source, which is denoted as $R_{X|Y}$ for comparison. We can observe that, by allowing the decoder to access the side information, the achievable rate under the same distortion and perception constraints can be reduced. We can also observe that, under the same achievable rate, increasing the perceptual quality at the decoder will lead to higher symbol-level distortion of the decoded signals.
	Furthermore, under the same distortion and perception constraints, the rate achieved by the indirect observation of semantic source information at the encoder is higher than that achieved when the semantic source information is fully observable by the encoder, i.e., $R_{S|Y}(D,P) > R_{X|Y}(D,P)$.
	When the side information is available at the decoder, even when the achievable rate becomes zero, i.e., $R(D, P)=0$, the symbol-level distortion, measured by MSE, as well as the perceptual quality of the recovered signal at the decoder will still be within a certain limit. This observation is consistent with that of binary information source discussed in Section \ref{Section_BinarySource}. In other words, even when the transmitter does not send any data to the channel, it is still possible for the decoder to recover some semantic source information within a certain distortion and perception constraints based on the side information.
	
	\section{Conclusions}
	This paper has investigated  the achievable data rate of semantic communication under the distortion and  perception constraints. We have considered a semantic communication model in which a semantic information source can only be indirectly observed by the encoder and the encoder and decoder can access to different types of side information.
	We derive the achievable rate region that characterizes the tradeoff among the data rate, symbol distortion, and semantic perception. We have then theoretically proved that
	there exists a stochastic coding scheme to achieve the derived region.
	We derive a closed-form achievable rate for binary semantic information source. We observe that there exist cases that the receiver can directly infer the semantic information source satisfying certain distortion and perception constraints without requiring any data communication from the transmitter. Experimental results based on the image semantic source signal have been presented to verify our theoretical observations.
	
	
	\section*{Acknowledgment}
	This work was supported in part by the National Natural Science Foundation of China under Grant 62071193 and the major key project of Peng Cheng Laboratory under grant PCL2023AS1-2.
	
	\appendices
	\section{Proof of Theorem \ref{th1}}
	\label{proof_th1}
	To prove the sufficiency part of the theorem,
	we adopt the source coding scheme using strong functional representation in \cite{theis2021coding} and generate a random codebook consisting of $2^{mR_1}$ i.i.d codewords where $R_1=I(S,X;Z|Y)-\epsilon$ for arbitrary small $\epsilon>0$. These codewords are then distributed uniformly randomly into $2^{kR_2}$ bins where $R_2=I(M;\hat M)-\epsilon$.
	Then we map $mR_1$-bit length codewords into a channel coding scheme with $kR_2$ bits.
	Since $kR_1\leq mR_2$ and $R_2$ correspond to the channel capacity, $S,X,Z,Y,M,\hat M$ are joint typical sequences.
	To prove the necessity part, we follow the same line as \cite{merhav2003joint} and prove that there exists a joint-source channel coding scheme that satisfies
	$
	I(S,X;\hat M)=I(S,X,Y;\hat M)\geq I(S,X;Z|Y)
	$
	Similarly, follow the same line as \cite{merhav2003joint}, we can prove that for any coding scheme including the proposed joint source channel coding scheme, the mutual information between the mutual information
	between the coded message and the channel output should not exceed the channel capacity $I(M;\hat M)$. Then we have $mI(S,X;\hat M)\leq kI(M;\hat M)$. This concludes the proof.
	
	\section{Proof of Theorem \ref{th2}}
	\label{proof_th2}
	Let us first define
	\begin{eqnarray}
		\label{cr0}
		R^{(p)}(D,P):=
		\left\{
		\begin{aligned}
			&R_{p}(D) \hspace{22pt}{\rm if}\ 0\leq D\leq D_{1,q}\ {\rm or}\ P\geq p\\
			&R_{p}(D,P) \quad {\rm if}\ D_{1,p}\leq D \leq D_{2,p}\\
			&0	\hspace{46pt} {\rm if}\ D\geq D_{2,p},
		\end{aligned}
		\right.
	\end{eqnarray}
	with
	$
	D_{1,p}=\frac{P}{1+2P-2p},
	D_{2,p}= 2p(1-p)-(1-2p)P,
	$
	and parameterize following conditional probabilities as
	$
	p_{\hat S|X}(0|0)=s,\ p_{\hat S|X}(0|1)=t,
	p_Y(0)=p_a,\ p_Y(1)=p_b.
	$
	Assume that $S$ and $X$ form the doubly symmetric binary distribution, then $p_S(1)=p_X(1)=\frac12$, thus we have $d(X,\hat S)=\frac12(1-s+t)$ and
	$
	p_{S\hat S}(0,1)=\frac12(1-q)(1-s)+\frac12 q(1-t)
	p_{S\hat S}(1,0)=\frac12 qs+\frac12(1-q)t.
	$
	The distortion between indirect semantic source and reconstruction signal is then formulated as
	$
	d(S,\hat S)=p_{S\hat S}(0,1)+p_{S\hat S}(1,0)
	=(1-2q)\cdot d(X,\hat S)+q,
	$
	which is further written as
	$
	\label{dhxs}
	d(X,\hat S)=\frac{d(S,\hat S)-q}{1-2q}.
	$
	Thus the indirect distortion is converted linearly as direct distortion.
	For perception constraint, as $p_X(1)=p_S(1)=\frac12$, we have $D_{TV}(p_S,p_{\hat S})=D_{TV}(p_X,p_{\hat S})$.
	We then have
	$
	d(X,\hat S)=p_a d_0+p_b d_1,
	$
	where $
	d_0=d(X,\hat S|Y=0),d_1=d(X,\hat S|Y=1)
	$.
	Similarly, the perception distortion can also be written as the expectation form based on following lemma:
	\begin{lemma}
		$
		D_{TV}(p_X,p_{\hat S})=|p_a\cdot p_0\pm p_b\cdot p_1|,
		$
		with
		$
		p_0 = D_{TV}(p_{X|Y=0},p_{\hat S|Y=0}),\ p_1 = D_{TV}(p_{X|Y=1},p_{\hat S|Y=1}).
		$
		If $D_{1,a}\leq d_0\leq D_{2,a},\ D_{1,b}\leq d_1\leq D_{2,b}$ and $p_0<a^*, p_1<b^*$, we have
		$
		D_{TV}(p_S,p_{\hat S})=p_a\cdot p_0+p_b\cdot p_1.
		$
	\end{lemma}
	We omit the proof due to limit of the space.
	Following (6) of \cite{blau2019rethinking}, we then reformulate the mutual information as
	$
		I(X;\hat S|Y)\geq\ p_a R^{(a^*)}(d_0,p_0)+p_b R^{(b^*)}(d_1,p_1).
	$
	Thus the optimization problem of minimizing the conditional mutual information of semantic source is written as
	\begin{eqnarray}
		\label{min2}
		&&\min_{d_0,d_1,p_0,p_1}\ p_a R^{(a^*)}(d_0,p_0)+p_b R^{(b^*)}(d_1,p_1) \\
		&&\;\;\; \mbox{s.t.} \;\;\;\quad  p_a((1-2q)d_0+q)+p_b((1-2q)d_1+q)	\leq D \nonumber \\
		&&\;\;\; \;\;\; \;\;\; \;\;\;  D_{TV}|p_a\cdot p_0\pm p_b\cdot p_1|\leq P \nonumber
	\end{eqnarray}
	To solve (\ref{min2}), we can associate a Lagrangian function $\mathcal{L}$ and the closed-form solution is obtained by solving the minimization problem of $\mathcal{L}$.
	The detailed proof is omitted.
	
	\bibliographystyle{IEEEtran}
	\bibliography{icnp}

\end{document}